\documentclass[12pt,sec]{article}
\usepackage{epsfig}
\usepackage{axodraw}
\begin{document}

\begin{center}
{\large  Production of $J/\psi +c\bar{c}$ through two photons
in $e^+e^-$ annihilation}\\[0.8cm]
{ Kui-Yong Liu  and Zhi-Guo He }\\[0.5cm]
{\footnotesize Department of Physics, Peking University,
 Beijing 100871, People's Republic of China}\\[0.5cm]

{Kuang-Ta Chao } \\[0.5cm]
{\footnotesize China Center of Advanced Science and Technology
(World Laboratory), Beijing 100080, People's Republic of China and
Department of Physics, Peking University,
 Beijing 100871, People's Republic of China}
\end{center}

\begin{abstract}

We study the production of $J/\psi+c\bar{c}$ in $e^+e^-$
annihilation through two virtual photons. The cross section is
estimated to be 23 fb at $\sqrt{s}=10.6$ GeV, which is smaller by
a factor of six than the calculated cross section for the same
process but through one virtual photon. As a result, while the
annihilation into two photons may be important for certain
exclusive production processes, the big gap between the inclusive
production cross section $\sigma(e^+e^- \rightarrow J/\psi +
c\bar{c})\simeq$ 0.9pb observed by Belle and the current
nonrelativistic QCD prediction of $\simeq$0.15pb still remains
very puzzling. We find, however, as the center-of-mass energy
increases ($\sqrt{s}>20$ GeV) the production through two virtual
photons $e^+e^- \rightarrow 2\gamma^* \rightarrow J/\psi +
c\bar{c}$ will prevail over that through one virtual photon,
because in the former process the photon fragmentation into
$J/\psi$ and into the charmed quark pair
becomes more important at higher energies.   \\

PACS number(s): 12.40.Nn, 13.85.Ni, 14.40.Gx

\end{abstract}

Charmonium production is one of the important processes to test
quantum chromodynamics (QCD) both perturbatively and
non-perturbatively. Because of the simpler parton structure
involved, which will reduce the theoretical uncertainty,
charmonium production in $e^+e^-$ annihilation is expected to be
more decisive in clarifying the production mechanisms of heavy
quarkonia. The two B factories with BaBar and Belle are collecting
huge data samples of continuum $e^+e^-$ annihilation events, which
will allow us to have a fine data analysis for charmonium
production. Recently, Charmonium production in $e^+e^-$
annihilation has become more interesting and puzzling, because of
the large gap between the Belle measurements\cite{belle1,belle2}
and the theoretical calculations for both
inclusice\cite{cho,yuan,baek,liu1} and exclusive
\cite{Braaten,liu2} charmonium production via double $c\bar{c}$
pairs based on nonrelativistic QCD (NRQCD).

For the inclusive processes, Belle has reported a measurement on
the $J/\psi$ production in $e^+e^-$ annihilation at
$\sqrt{s}=10.6$ GeV\cite{belle1,belle2}, and found that a very
large fraction of the produced $J/\psi$ is due to the double
$c\bar{c}$ production in $e^+e^-$ annihilation\cite{belle1}

\begin{equation}
\sigma(e^+e^- \rightarrow J/\psi c\bar{c})/\sigma(e^+e^-
\rightarrow J/\psi X)=0.59^{+0.15}_{-0.13}\pm0.12,
\end{equation}
which corresponds to \cite{belle1,belle2}
\begin{equation}
\sigma(e^+e^- \rightarrow J/\psi c\bar{c})\approx0.9 pb.
\end{equation}

In contrast, the predicted values for the cross section in NRQCD
(the color-octet contribution is negligible for this process) are
much smaller than the data\cite{cho,yuan,baek}. In a recent
analytical calculation and numerical estimation for the inclusive
charmonia production including all S-wave, P-wave and D-wave
states via double $c\bar{c}$ in NRQCD\cite{liu2}, we find
\begin{equation}
\sigma(e^+e^- \rightarrow \gamma^* \rightarrow J/\psi
c\bar{c})\approx 0.15 pb.
\end{equation}
This value is consistent with other previous calculations,
including those obtained based on the quark-hadron duality
hypothesis\cite{kiselev}, but smaller than the Belle data by about
a factor of six\footnote{The numerical value obtained for this
process in\cite{cho} should be multiplied by a factor of 3.}.

For the exclusive processes, the Belle measurement\cite{belle1}
for the cross section of $e^+e^- \rightarrow J/\psi + \eta_c$ is
about an order of magnitude larger than the NRQCD calculation for
$e^+e^- \rightarrow \gamma^* \rightarrow J/\psi +
\eta_c$\cite{Braaten,liu2}. In order to solve the problem,
calculations for the exclusive double-charmonium production from
$e^+e^-$ annihilation into two virtual photons are
performed\cite{bodwin}, and it is pointed out that the cross
section for $e^+e^- \rightarrow 2\gamma^* \rightarrow J/\psi +
J/\psi$ is larger than that for $e^+e^- \rightarrow \gamma^*
\rightarrow J/\psi + \eta_c$ by about a factor of 3.7, despite of
possible uncertainties due to the choice of input
parameters\cite{luchinsky}. This is a interesting result since it
indicates that the $e^+e^-$ annihilation through two virtual
photon fragmentation may make important contributions to certain
processes, and it might substantially reduce the discrepancy
between experiment and theory for the exclusive process $e^+e^-
\rightarrow J/\psi + \eta_c$ (note that it is essential to check
experimentally whether $e^+e^- \rightarrow J/\psi + J/\psi$ is
largely misidentified with $e^+e^- \rightarrow J/\psi + \eta_c$).

In this situation, it is necessary to examine the contribution of
$e^+e^-$ annihilation through two virtual photons to the inclusive
production of $J/\psi$. In the following we will calculate the
complete {\cal O}$(\alpha^4)$ color-singlet inclusive cross
section for $e^+e^- \rightarrow 2\gamma^* \rightarrow J/\psi
c\bar{c}$, and compare the production rate through two virtual
photons with that through one virtual photon, to see whether the
annihilation through two virtual photons can decrease the
discrepancy between the Belle measurement on $J/\psi c\bar{c}$
production and the calculations based on NRQCD.

\begin{figure}[t]
\begin{center}
\vspace{-2.8cm}
\includegraphics[width=14cm,height=16cm]{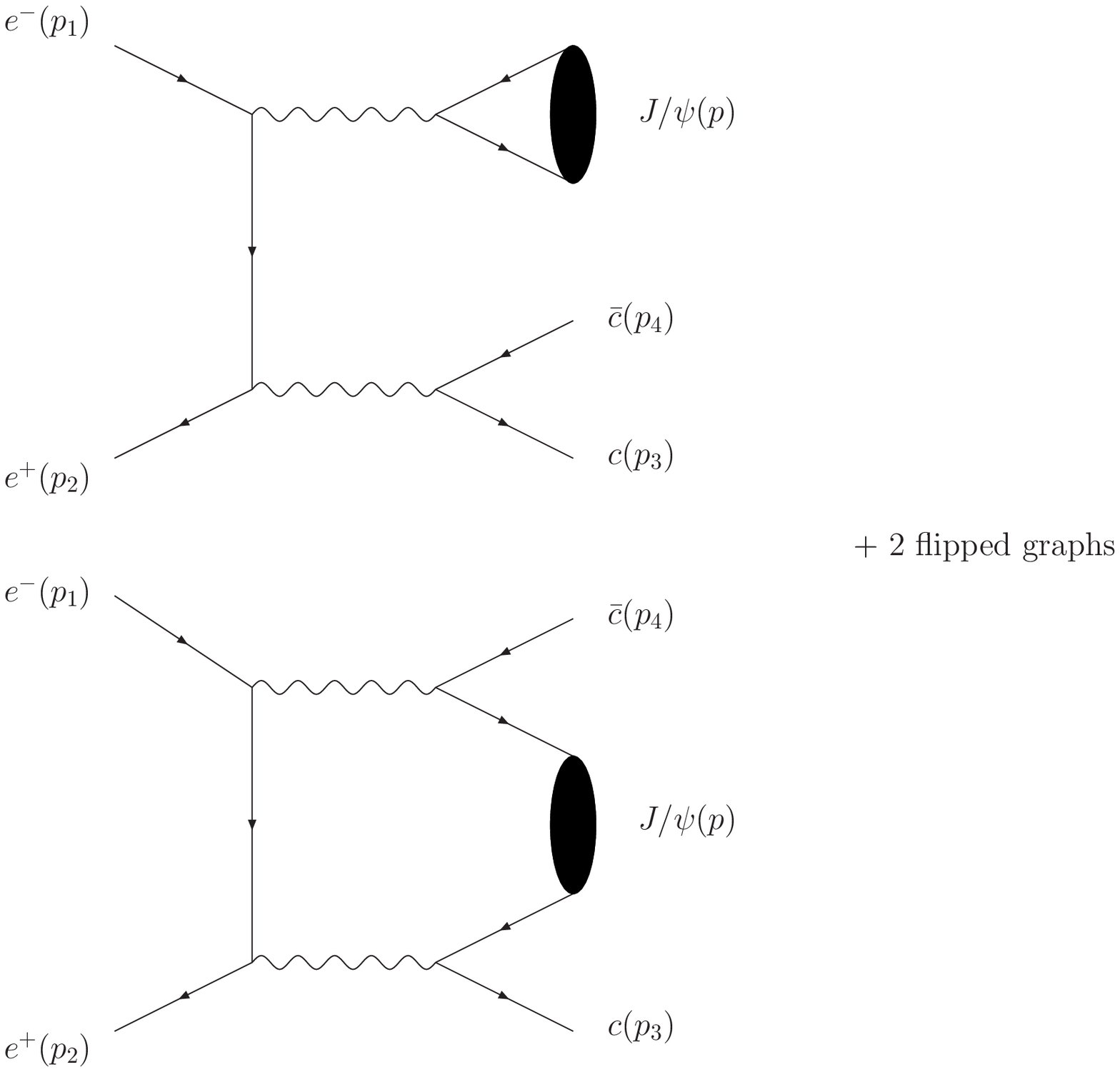}
\vspace{-3cm}
\end{center}
\caption{ Feynman diagrams for $e^+e^-\rightarrow 2 \gamma^*
\rightarrow J/\psi+c\bar{c} $.} \label{feynman}
\end{figure}

Following the NRQCD factorization formalism, color-singlet
scattering amplitude of the process $e^-(p_1)+e^+(p_2)\rightarrow
\gamma^* \rightarrow c\bar{c}(^{2S+1}L_J)(p)+c(p_3)+\bar{c}(p_4)$
in Fig.~\ref{feynman} is given by
\begin{eqnarray}
%\hspace{-1.0cm}\hspace{1.0cm}
\label{amp2}   &&\hspace{-1.5cm} {\cal
A}(e^-(p_1)+e^+(p_2)\rightarrow
c\bar{c}(^{2S+1}L_{J})(p)+c(p_3)+\bar{c}(p_4))=\sqrt{C_{L}}
\sum\limits_{L_{z} S_{z} }\sum\limits_{s_1s_2 }\sum\limits_{jk}
\nonumber\\
&&\hspace{-1.5cm}\times\langle s_1;s_2\mid S S_{z}\rangle \langle
L
L_{z};S S_{z}\mid J J_{z}\rangle\langle 3j;\bar{3}k\mid 1\rangle\nonumber\\
&&\hspace{-1.5cm}\times{\cal A}(e^-(p_1)+e^+(p_2)\rightarrow
 c_j(\frac{p}{2};s_1)+\bar{c}_k(\frac{p}{2};s_2)+
 c_l(\frac{p_3}{2};s_3)+\bar{c}_i(\frac{p_4}{2};s_4)) (L=S).
\end{eqnarray}
where $c\bar{c}(^{2S+1}L_{J})$ is the intermediate $c\bar{c}$ pair
which is produced at short distances and then evolves into a
specific charmonium at long distances. $\langle 3j;\bar{3}k\mid
1\rangle =\delta_{jk}/\sqrt{N_c}$~, ~ $\langle s_1;s_2\mid S
S_{z}\rangle$~,~$ \langle L L_z ;S S_z\mid J J_z\rangle$ are
respectively the color-SU(3), spin-SU(2), and angular momentum
Clebsch-Gordon coefficients for $Q\bar{Q}$ pairs projecting out
appropriate bound states. For $J/\psi$ the coefficient $C_{L}$ can
be related to the radial wave function of the bound state and
reads
\begin{equation}
C_{S}=\frac{1}{4\pi}\mid R_{S}(0) \mid^{2}.
\end{equation}

The spin projection operator can be defined as\cite{pro}
\begin{equation}
P_{SS_z}(p;q)\equiv\sum\limits_{s_1s_2 }\langle
s_1;s_2|SS_z\rangle
v(\frac{p}{2}+q;s_1)\bar{u}(\frac{p}{2}-q;s_2).
\end{equation}

We write the spin projection operator which will be used in the
calculation as

\begin{equation}
P_{1S_Z}(p,0)=\frac{1}{2\sqrt{2}}\not{\epsilon}(S_z)(\not{p}+M),
\end{equation}
where $M$ is the mass of the charmonium, which equals to $2m_c$ in
the nonrelativistic approximation.

\begin{figure}[t]
\begin{center}
\vspace{-1cm}
\includegraphics[width=12cm,height=10cm]{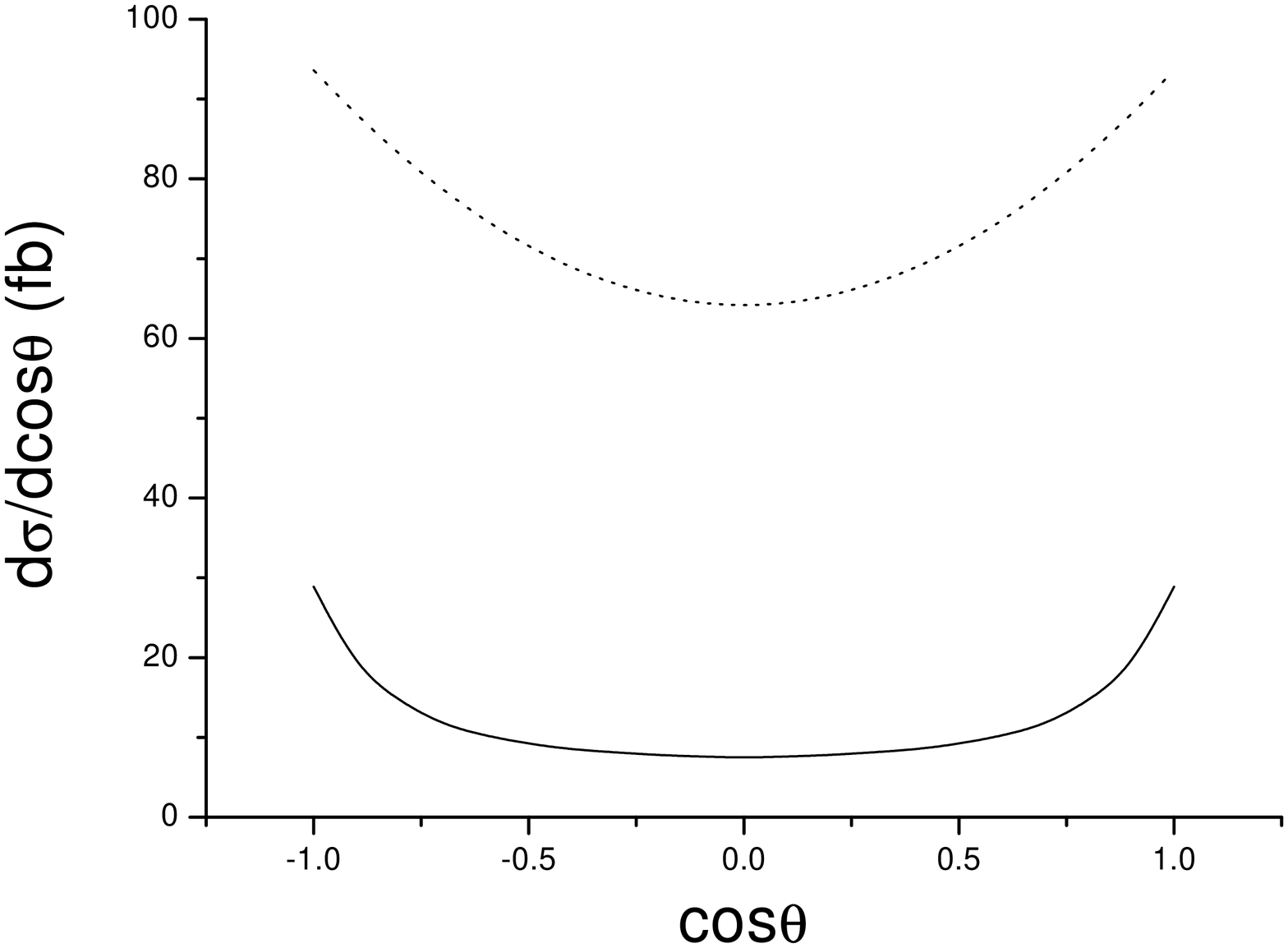}
\vspace{-1cm}
\end{center}
\caption{Differential cross sections of $e^+e^-\rightarrow
2\gamma^* \rightarrow J/\psi + c\bar{c}$ (solid line) and
$e^+e^-\rightarrow \gamma^* \rightarrow J/\psi + c\bar{c}$ (dotted
line) as functions of the scattering angle of $J/\psi$.}
\label{distribution}
\end{figure}

The amplitude for the upper Feynman graph in Fig.~\ref{feynman}
can be written as
\begin{eqnarray}
M_1&=&\frac{4\sqrt{3}}{9}\delta_{kl}\sum_{L_zS_z}\langle L L_{z};S
S_{z}\mid J J_{z}\rangle
\bar{v}(p_2)\gamma_{\mu}\frac{\not{p_1}-\not{p}+m_e}{(p_1-p)^2-m_e^2}
\gamma_{\alpha}u(p_1) \nonumber \\
&&Tr[P_{1S_z}(p,0)\gamma_{\alpha}]\bar{u}(p_3)\gamma_{\mu}v(p_4)
\frac{1}{p^2(p_3+p_4)^2},
\end{eqnarray}
and the amplitude for the corresponding flipped graph is denoted
as $M_2$. The amplitude for the lower Feynman graph in
Fig.~\ref{feynman} reads
\begin{eqnarray}
M_3&=&\frac{4}{9\sqrt{3}}\delta_{kl}\sum_{L_zS_z}\langle L L_{z};S
S_{z}\mid J J_{z}\rangle
\bar{v}(p_2)\gamma_{\mu}\frac{\not{p_3}+\not{p}/2-\not{p_2}+m_e}{(p_3+p/2-p_2)^2-m_e^2}
\gamma_{\alpha}u(p_1) \nonumber \\
&&\bar{u}(p_3)\gamma_{\mu}P_{1S_z}(p,0)\gamma_{\alpha}v(p_4)\frac{1}{(p_4+p/2)^2}\frac{1}{(p_3+p/2)^2},
\end{eqnarray}
and the amplitude for the flipped graph is denoted as $M_4$.

The calculation of cross section for $e^{-}+e^{+}\rightarrow
2\gamma^*\rightarrow J/\psi + c\bar{c}$ is straightforward. The
differential cross section can be written in the form
\begin{equation}
\label{cross} \frac{d\sigma(e^{+}+e^{-}\rightarrow 2\gamma^{*}
\rightarrow J\psi +
c\bar{c})}{d\Omega}=\frac{3C_{S}\alpha^{4}}{4m_{c}}\mid \bar{M}
\mid^2,
\end{equation}
where $d\Omega$ represents the elements of the quadruple integral
related to the final state phase space (see, e.g.
Ref.~\cite{cho}), and $\mid \bar{M}
\mid^2=\frac{1}{4}\sum_{pol,color}\mid M_1+M_2+M_3+M_4 \mid^2$ is
the unpolarized module squared of the amplitude. For simplicity we
will not write down their lengthy expressions here.

With Eq.~(\ref{cross}) we can evaluate the inclusive cross
sections for $J/\psi$ production from $e^+e^-$ through two virtual
photons. The input parameters used in the numerical calculations
are the same as in ref.\cite{liu1}
\begin{equation}
m_e=0, ~~m_c=1.5GeV, ~~\alpha=1/137,
\end{equation}
\begin{equation}
\mid R_S(0) \mid^2=0.81GeV^3\cite{wf}.
\end{equation}

Now we give the numerical result at the Belle energy
$\sqrt{s}=10.6$ GeV:

\begin{equation}
\label{js} \sigma(e^{+}+e^{-}\rightarrow 2\gamma^{*}\rightarrow
J/\psi+c\bar{c})=23 {\rm fb}.
\end{equation}

\begin{figure}[t]
\begin{center}
\vspace{5cm}
\includegraphics[width=12cm,height=10cm]{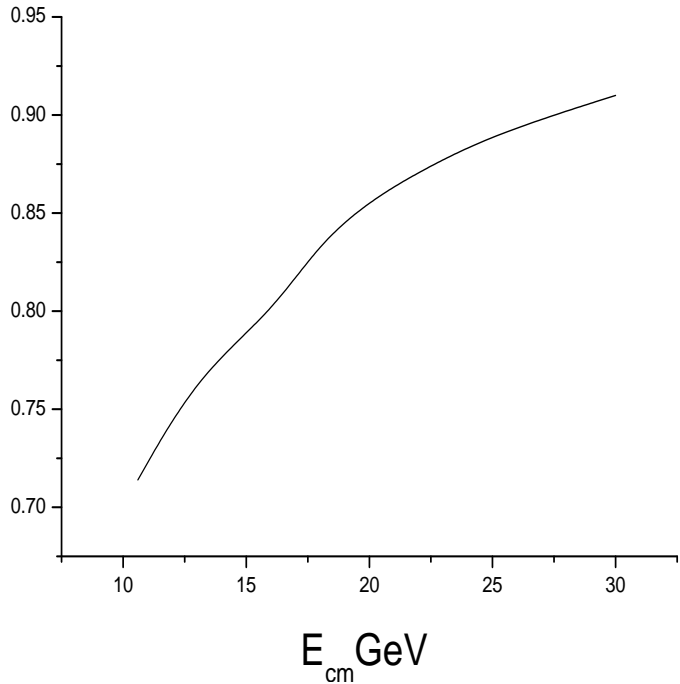}
\vspace{-5cm}
\end{center}
\caption{ Ratio of the fragmentation contribution to the total
cross section as the function of the center-of-mass energy.}
\label{ratio}
\end{figure}

It is well known that $e^+e^- \rightarrow 2\gamma^* \rightarrow
J/\psi + c\bar{c}$ is a pure electromagnetic process (except for
the hadronization of the quark pair at long distances), while
$e^+e^- \rightarrow \gamma^* \rightarrow J/\psi + c\bar{c}$
involves both electromagnetic and strong interactions. For a naive
order of magnitude estimate, the ratio of the production rate of
the former to the latter would be proportional to
$\alpha^2/\alpha_s^2$, but the photon fragmentation into $J/\psi$
and into the charmed quark pair can substantially enhance the
former. To see the role of photon fragmentation, we plot the
differential cross sections of $e^+e^-\rightarrow 2\gamma^*
\rightarrow J/\psi + c\bar{c}$ and $e^+e^-\rightarrow \gamma^*
\rightarrow J/\psi + c\bar{c}$ as functions of the scattering
angle of $J/\psi$ at $\sqrt{s}=10.6$GeV in
Fig.~\ref{distribution}. (Here for the one-photon process we
choose $\alpha_s=$0.26.) One can see that the small angle $J/\psi$
production in $e^+e^-\rightarrow 2\gamma^* \rightarrow J/\psi +
c\bar{c}$ is dominant. This indicates that most of the $J/\psi$
production comes from the photon fragmentation (corresponding to
the upper graph in Fig.~\ref{feynman} with one photon fragmenting
to $J/\psi$ and the other fragmenting to a charm quark pair).
Indeed, our calculation shows that at $\sqrt{s}=10.6$GeV, the
contribution from fragmentation graphes is about seventy-two
percent of the total cross section. In Fig.~\ref{ratio} we show
the ratio of the fragmentation contribution to the total cross
section as a function of the $e^+e^-$ center-of-mass energy in
$e^+e^-\rightarrow 2\gamma^* \rightarrow J/\psi + c\bar{c}$. It is
clear that the photon fragmentation becomes more and more dominant
as the center-of-mass energy increases. This is in agreement with
the observation in the $J/\psi + J/\psi$ exclusive production
through two virtual photons\cite{bodwin}.

\begin{figure}[t]
\begin{center}
\vspace{5cm}
\includegraphics[width=12cm,height=10cm]{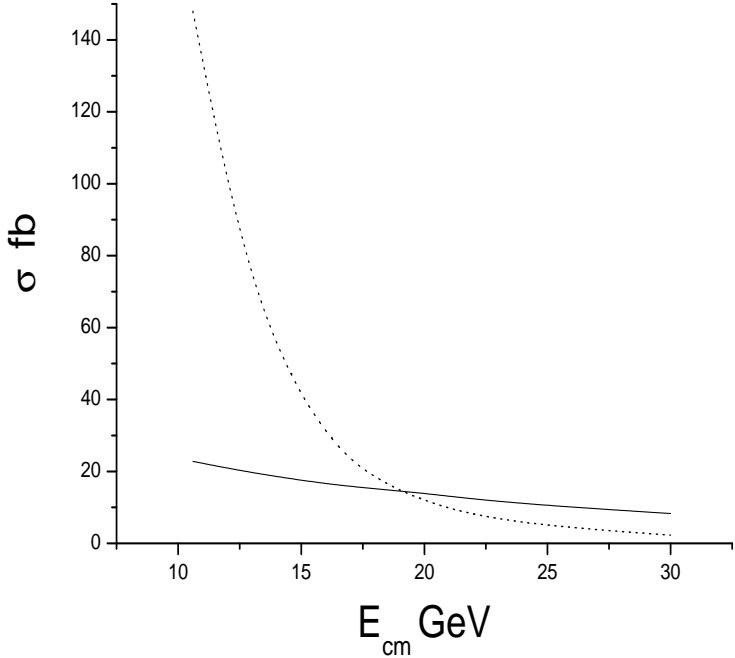}
\vspace{-5cm}
\end{center}
\caption{Cross sections of $e^+e^-\rightarrow 2\gamma^*
\rightarrow J/\psi + c\bar{c}$ (solid line) and $e^+e^-\rightarrow
\gamma^* \rightarrow J/\psi + c\bar{c}$ (dotted line) as functions
of the center-of-mass energy} \label{num}
\end{figure}

From the above discussions we have seen the importance of the
photon fragmentation to $J/\psi$ as well as to the charm quark
pair in the two-photon process $e^+e^-\rightarrow 2\gamma^*
\rightarrow J/\psi + c\bar{c}$. A even more crucial result is that
at high $e^+e^-$ center-of-mass energies the contribution through
two virtual photons will prevail over that through one virtual
photon in the production of $J/\psi c\bar{c}$. The reason lies
simply in the fact that the virtuality of the photon in the
two-photon process can be as small as $4m_c^2$, whereas it is as
large as $s$, the center-of-mass energy squared in the one-photon
process. In Fig.~\ref{num} we show the cross sections of
$e^+e^-\rightarrow 2\gamma^* \rightarrow J/\psi + c\bar{c}$ (solid
line) and $e^+e^-\rightarrow \gamma^* \rightarrow J/\psi +
c\bar{c}$ (dotted line) as functions of the center-of-mass energy
$\sqrt{s}$. We see clearly that the cross section for
$e^+e^-\rightarrow 2\gamma^* \rightarrow J/\psi + c\bar{c}$
decreases very slowly, whereas that for $e^+e^-\rightarrow
\gamma^* \rightarrow J/\psi + c\bar{c}$ decreases rapidly as
$\sqrt{s}$ increases. At $\sqrt{s}=20$ GeV, the two photon process
becomes to prevail over the one photon process.

However, unfortunately, at the Belle energy $\sqrt{s}=10.6$ GeV,
since the enhancement effect due to the factor $s/m_c^2$ is not
large enough as compared with the suppression factor
$\alpha^2/\alpha_s^2$, we find $\sigma(e^+e^-\rightarrow 2\gamma^*
\rightarrow J/\psi + c\bar{c})=23$ fb, which is still much smaller
than $\sigma(e^+e^-\rightarrow \gamma^* \rightarrow J/\psi +
c\bar{c})=148$ fb\cite{liu1}, and therefore is negligible.

In summary, we have calculated the complete ${\cal O}(\alpha^{4})$
color-singlet inclusive cross sections for $J/\psi c\bar{c}$
production from $e^+ e^-$ annihilation into two photons. Due to
the suppression factor of $\alpha^2/\alpha_s^2$, at the $e^+e^-$
center-of-mass energy $\sqrt{s}=10.6$ GeV the cross section of
this process is smaller by about a factor of six than that from
$e^+ e^-$ annihilation into one photon. We then conclude that
while the $e^+ e^-$ annihilation into two photons could be helpful
in solving the puzzle for the exclusive $J/\psi\eta_c$ production,
it can do very little to reduce the big gap between the observed
inclusive production cross section of $\sigma(e^+e^- \rightarrow
J/\psi + c\bar{c})\approx 0.9$ pb and the current NRQCD
predictions of about 0.15 pb. This puzzle still needs to be
explained with new theoretical considerations. We find, however,
as the center-of-mass energy increases($\sqrt{s}>$20 GeV) the
production through two photons $e^+e^-\rightarrow 2\gamma^*
\rightarrow J/\psi + c\bar{c}$ will prevail over that through one
photon, because in the former case the photon fragmentation into
$J/\psi$ and into the charmed quark pair becomes more important at
higher energies.

\section*{Acknowledgments}

This work was supported in part by the National Natural Science
Foundation of China, and the Education Ministry of China.

\end{document}